\documentclass[pdflatex,sn-mathphys]{sn-jnl}% Math and Physical Sciences Reference Style
\DeclareUnicodeCharacter{2113}{L}
\jyear{2021}%

\theoremstyle{thmstyleone}%
\theoremstyle{thmstyletwo}%

\theoremstyle{thmstylethree}%

\begin{document}

\title[Deep transcoding of EEG/fMRI]{Inferring latent neural sources via  deep transcoding of simultaneously acquired EEG and fMRI}


\author[1]{\fnm{Xueqing} \sur{Liu}}\email{xl2556@columbia.edu}

\author[1]{\fnm{Tao} \sur{Tu}}\email{tt2531@columbia.edu}
\author[1,2,3,4]{\fnm{Paul} \sur{Sajda}}\email{psajda@columbia.edu}

\affil[1]{\orgdiv{Department of Biomedical Engineering}, \orgname{Columbia University}, \postcode{10027}, \state{NY}, \country{USA}}

\affil[2]{\orgdiv{Department of Electrical Engineering}, \orgname{Columbia University}, \postcode{10027}, \state{NY}, \country{USA}}

\affil[3]{\orgdiv{Department of Radiology}, \orgname{Columbia University}, \postcode{10031}, \state{NY}, \country{USA}}

\affil[4]{\orgdiv{Data Science Institute}, \orgname{Columbia University}, \postcode{10027}, \state{NY}, \country{USA}}


\abstract{Simultaneous EEG-fMRI is a multi-modal neuroimaging technique that provides complementary spatial and temporal resolution. Challenging has been developing principled and interpretable approaches for fusing the modalities, specifically approaches enabling inference of latent source spaces representative of neural activity. In this paper, we address this inference problem within the framework of  transcoding -- mapping from a specific encoding (modality) to a decoding (the latent source space) and then encoding the latent source space to the other modality. Specifically, we develop a symmetric  method consisting of a cyclic convolutional transcoder that transcodes EEG to fMRI and vice versa. Without any prior knowledge of either the hemodynamic response function or lead field matrix, the complete data-driven method exploits the temporal and spatial relationships between the modalities and latent source spaces to learn these mappings. We quantify, for both the simulated and real EEG-fMRI data, how well the modalities can be transcoded from one to another as well as the source spaces that are recovered, all evaluated on unseen data.  In addition to enabling a new way to symmetrically infer a latent source space, the method can also be seen as low-cost computational neuroimaging -- i.e. generating an 'expensive' fMRI BOLD image from 'low cost' EEG data. }


\maketitle

\section{Introduction}\label{sec1}

Functional magnetic resonance imaging (fMRI) is a neuroimaging modality that is a workhorse for cognitive neuroscience and increasingly used by clinical psychiatric departments \cite{glover2011overview}. fMRI has the advantage of full-brain coverage at relatively high spatial resolution (millimeters), though its temporal resolution is somewhat limited due to the sluggishness of the hemodynamic response \cite{huettel2004functional}. On the other hand, electroencephalography (EEG) is a neuroimaging modality with high temporal resolution (milliseconds) and low spatial resolution as it records electrical signals from electrodes on the surface of the scalp \cite{niedermeyer2005electroencephalography}. In light of such complementarity between the two modalities, when acquired simultaneously, EEG and fMRI potentially can compensate for the shortcoming in one modality using the merits of the other. 

An active area has been the development of machine learning approaches for fusing EEG and fMRI \cite{conroyfusing, oberlin2015symmetrical, transcoding, philiastides2021}. One major challenge, however, stems from the fact that EEG and fMRI capture distinct aspects of the underlying neuronal activity, and thus inevitably provide biased and only partially overlapping representations of the latent neural sources  \cite{jorge2014eeg}. An optimal cross-modal fusion, should therefore be capable of 1) identifying the overlapping neuronal substrates for both modalities while minimizing modality-specific bias, and 2) extracting and leveraging any potential information not shared by the two modalities but recorded by either modality, to optimize fusion.

EEG-informed-fMRI 
 \cite{benar2007single, jann2009bold, walz2013simultaneous, muraskin2016brain, muraskin2017fusing, muraskin2018multimodal} 
and fMRI-informed EEG \cite{debener2005trial} are the two  approaches for fusing simultaneously acquired EEG-fMRI. They are asymmetrical methods \cite{philiastides2021} in that only partial information from one modality is used to inform analysis of the other \cite{jorge2014eeg}. EEG-informed fMRI uses features from the EEG to build input regressors in voxel-wise fMRI general linear model (GLM) analyses \cite{jorge2014eeg}. For instance, EEG features can be extracted from trial-to-trial event related potentials \cite{benar2007single}, source dipole time series \cite{muraskin2016brain}, global EEG synchronization in the alpha frequency band \cite{jann2009bold}, or single-trial EEG correlates of task related activity \cite{walz2013simultaneous,muraskin2017fusing,muraskin2018multimodal}. These features are then convolved with a canonical hemodynamic response functions (HRF) before input to a GLM. The canonical HRF normally peaks around 4 to 5 seconds peristimulus time, lasts 20 to 30 seconds and changes very slowly. It is at best a rough estimate of the hemodynamic coupling with the underlying neural activity, and there is substantial research reporting significant variance in the true HRF, between subjects as well as within a subject across different brain regions \cite{handwerker2012continuing}.  On the other hand, fMRI-informed EEG analyses apply methods such as fMRI-informed source modeling\cite{debener2005trial} to constrain EEG source localization with a spatial prior provided by information from fMRI. EEG source localization often needs to employ a very complex model of the electromagentic field distribution of the head to calculate the forward and inverse models needed to estimate the location of neural sources in 3D space given the channel recordings on the scalp. These forward and inverse models are based on a leadfield matrix which is typically estimated from tissue conductivity and requires complex electromagnetic simulations.

Symmetrical methods have been developed which treat EEG and fMRI in a more balanced way \cite{philiastides2021}. For instance, Conroy, et al. \cite{conroyfusing} transformed both EEG and fMRI into the same data space through Canonical Correlation Analysis (CCA). Another example of symmetrical approaches \cite{oberlin2015symmetrical} maps the data fusion problem into an optimization problem, but this approaches still rely on an accurate estimate of the HRF and leadfield matrix, and cannot handle possible non-linearities that may exist between the EEG and fMRI data.

In this paper, we use simultaneously acquired EEG-fMRI and a novel convolutional neural network (CNN) structure to learn the  relationship between EEG and fMRI, and vice versa. Building on a previous work  \cite{transcoding} where a transcoder was developed and tested on simulated data, we substantially develop the approach and leverage the concept of Cycle-Consistent Adversarial Networks (CycleGAN) \cite{zhu2017unpaired}, to create a "cyclic-CNN" for transcoding of simultaneously acquired EEG and fMRI data from actual EEG-fMRI experiments. The results show that 1) our model can reconstruct fMRI data from EEG data, and vice versa, without any prior knowledge of hemodynamic coupling and leadfield estimates; 2) our model can  accurately estimate the underlying HRF and forward and inverse head models, without prior knowledge of the tissue conductivity or the need for complex electromagnetic simulations; 3) the model can also reveal the dynamics of the latent source space, enabling new ways of assessing the underlying network structure that accounts for the observed EEG and fMRI data.

\section{Results}

\begin{figure}[thpb]
  \centering
  \includegraphics[width=\columnwidth]{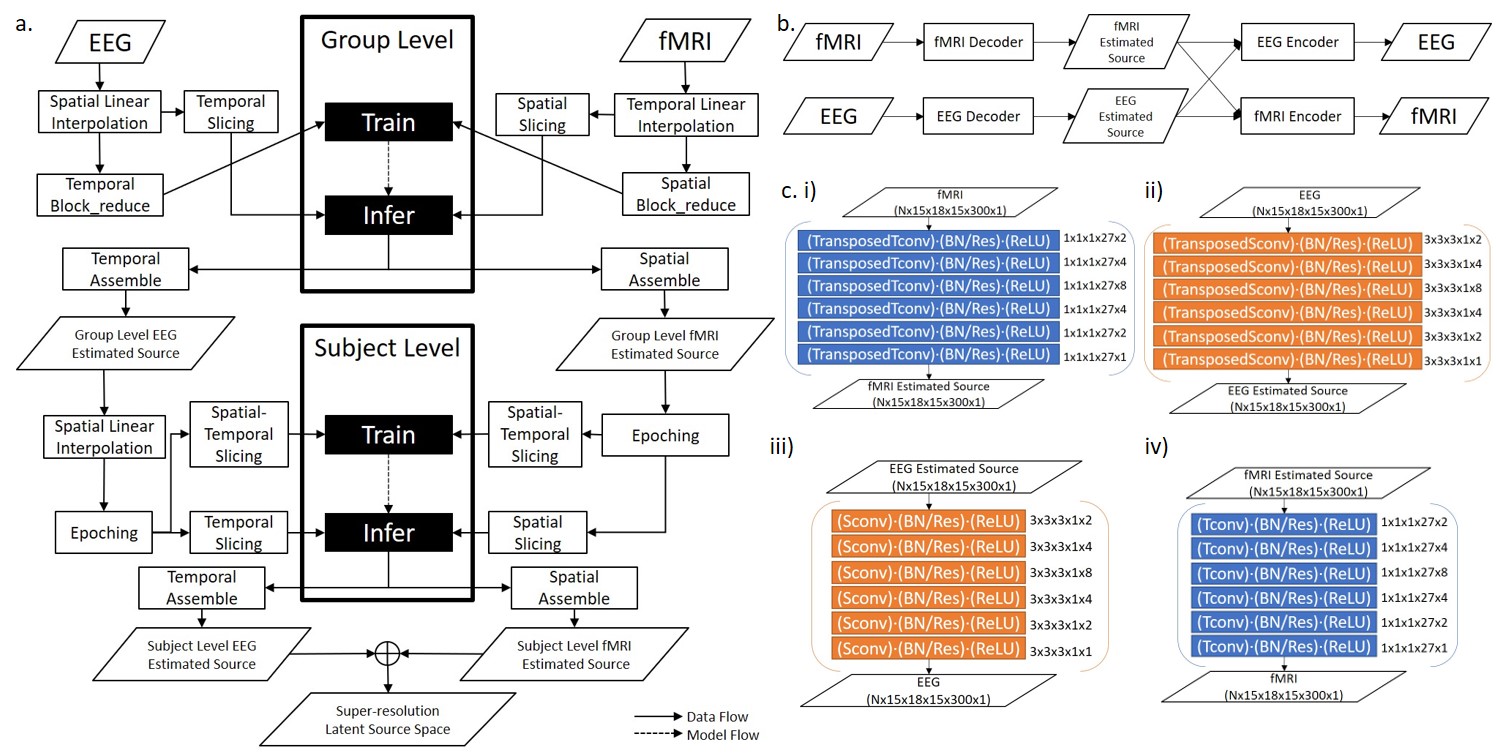}
  \caption{\textbf{a.} Framework of hierarchical deep transcoding for fusing simultaneous EEG/fMRI: Both the group level model and subject level model are cyclic convolutional transcoders as shown in b. 
  \textbf{b.} Framework of a cyclic convolutional transcoder: The transcoder is  made of an fMRI decoder, an EEG encoder, an EEG decoder and an fMRI encoder. 
  \textbf{c.} Detailed structure of the i) fMRI decoder, ii) EEG decoder, iii) EEG encoder, iv) fMRI encoder. In group level transcoders, batch-normalization layers are used while in subject level transcoders, residual layers are used instead.}
  \label{Method}
\end{figure}

\subsection{Transcoder Model}

Fig.~\ref{Method}a illustrates the overall pipeline of the hierarchical deep transcoding model. The model is composed of two stages (i.e. the hierarchy). In the first stage, the model is trained at the group level, on all of the data across the subject population,  to reach a intermediate spatiotemporal resolution. In the second stage, a subject level model is trained for each subject's individual data to finally reach the millimeter/millisecond resolution of the latent source space. The cyclic convolutional transcoder, as shown in Fig.~\ref{Method}b, is the core of the hierarchical deep transcoding structure. To briefly review the concept of "neural" transcoding, the idea is to generate a signal of one neuroimaging modality from another by first decoding it into a latent source space and then encoding it into the other measurement space.  Both the group level model and the subject level model take the shape of a cyclic convolutional transcoder.  The cyclic convolutional transcoder is made of four modules, namely an fMRI decoder, fMRI encoder, EEG decoder and EEG encoder. Decoders generate the latent source space from an encoding (EEG/fMRI), while an encoder maps the latent source space into an the measurement space (EEG or fMRI). Additional details of the model and how it is trained are provided in the Methods.

\subsection{Recovering sources in real simultaneous EEG-fMRI datasets}\label{real data results}
   
\subsubsection{Auditory oddball task}
\begin{figure}[thpb]
  \centering
  \includegraphics[scale=0.35]{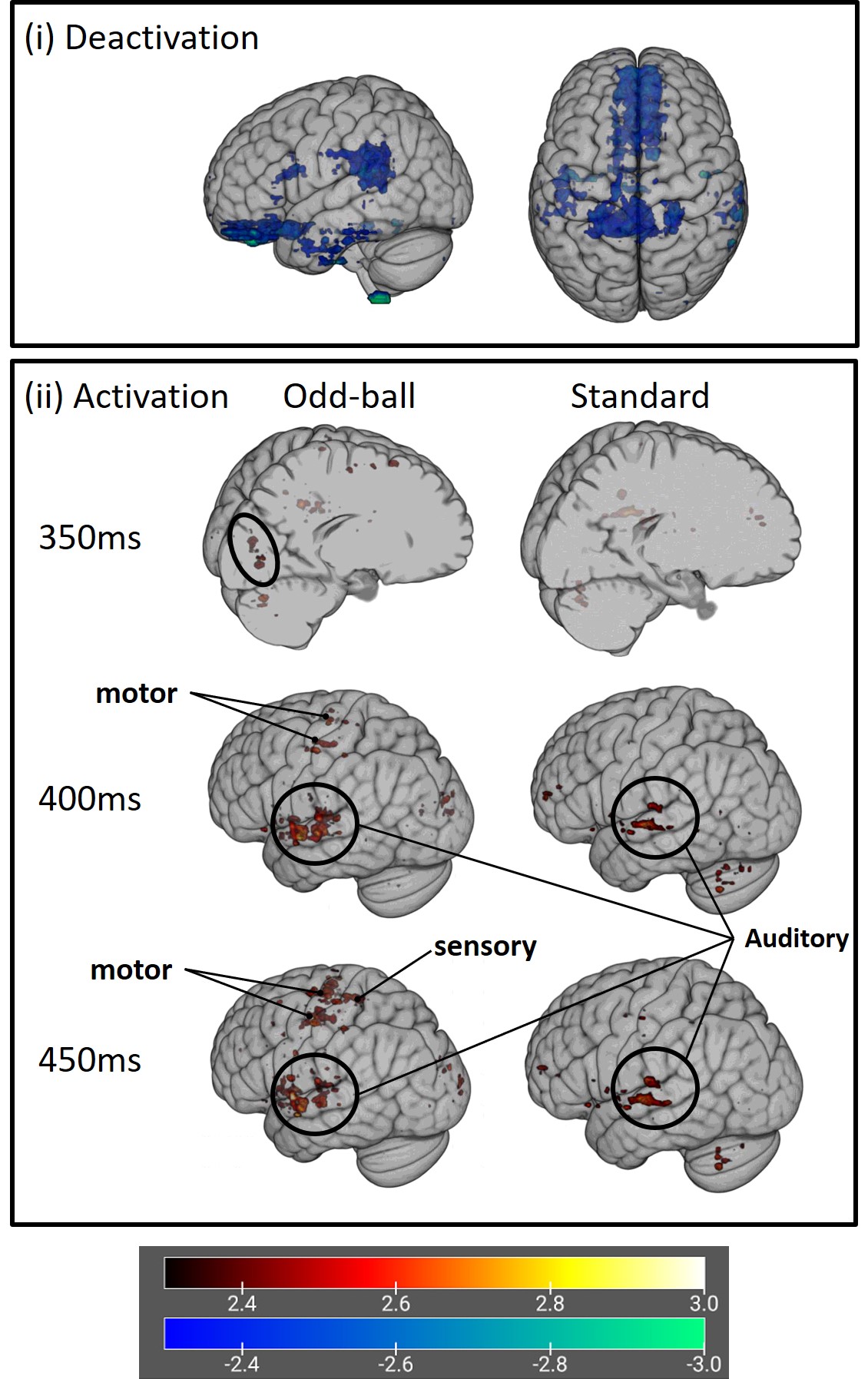}
  \caption{Representative time frames of  deactivation (i)  and activation (ii) in the spatio-temporal latent source space  (uncorrected Z-maps) of Auditory Odd-ball dataset.}
  \label{AuditoryOddBallCompactResult}
\end{figure}

Significant prolonged deactivation are present in both standard and oddball trials. The difference in deactivation patterns between the standard and oddball trials are not significant. Fig.\ref{AuditoryOddBallCompactResult}i shows the deactivation in one representative time frame of standard trials at 350 ms post-stimulus. The deactivations span across prefrontal cortex, posterior cingulate cortex, and temporal pole, consistent with previous findings  \cite{wolf2008auditory}. These regions form a network commonly identified as the default mode network (DMN)  which has been reported to deactivate when participants had to perform external goal-directed tasks \cite{fox2005human}. The deactivation of DMN was also reported by a previous simultaneous EEG-fMRI visual oddball study \cite{walz2014simultaneous} where the fMRI activity was correlated negatively with EEG single-trial variability (STV) at 525ms stimulus-locked window. A major advantage of our method over the asymmetrical EEG-fMRI fusion method (e.g. \cite{walz2014simultaneous}) is that no assumption needs to be made about the task-specific subspaces. The symmetrical data-driven nature of our method can therefore provide less biased estimates of the underlying source representations shared between EEG and fMRI.

In line with previous ECoG findings \cite{lecaignard2021dynamics}, prolonged activations were detected in the auditory cortex in both standard and oddball trials as shown in Fig.\ref{AuditoryOddBallCompactResult}ii. We also observed positive activations in the occipital lobe at 350 ms post-stimulus for oddball stimuli. This activity was absent following the onset of standard stimuli. It is associated with the well-studied stimulus evoked response P300 during oddball task \cite{polich1995cognitive}.

Moreover, around 400 ms (near mean response time) after the onset of the oddball stimuli, significant activations were observed in the left primary motor cortex which is likely related to subjects  motor response to oddball trials, as shown in Fig.\ref{AuditoryOddBallCompactResult}ii. At a slightly later time window (around 450 ms), positive activations extended to the primary somatosensory cortex which is likely related to sensory information processing with input from the right hand and index finger after the subject pressed the button. No such activation was observed after the onset of standard stimuli. 

\begin{figure}[thpb]
  \centering
  \includegraphics[scale=0.35]{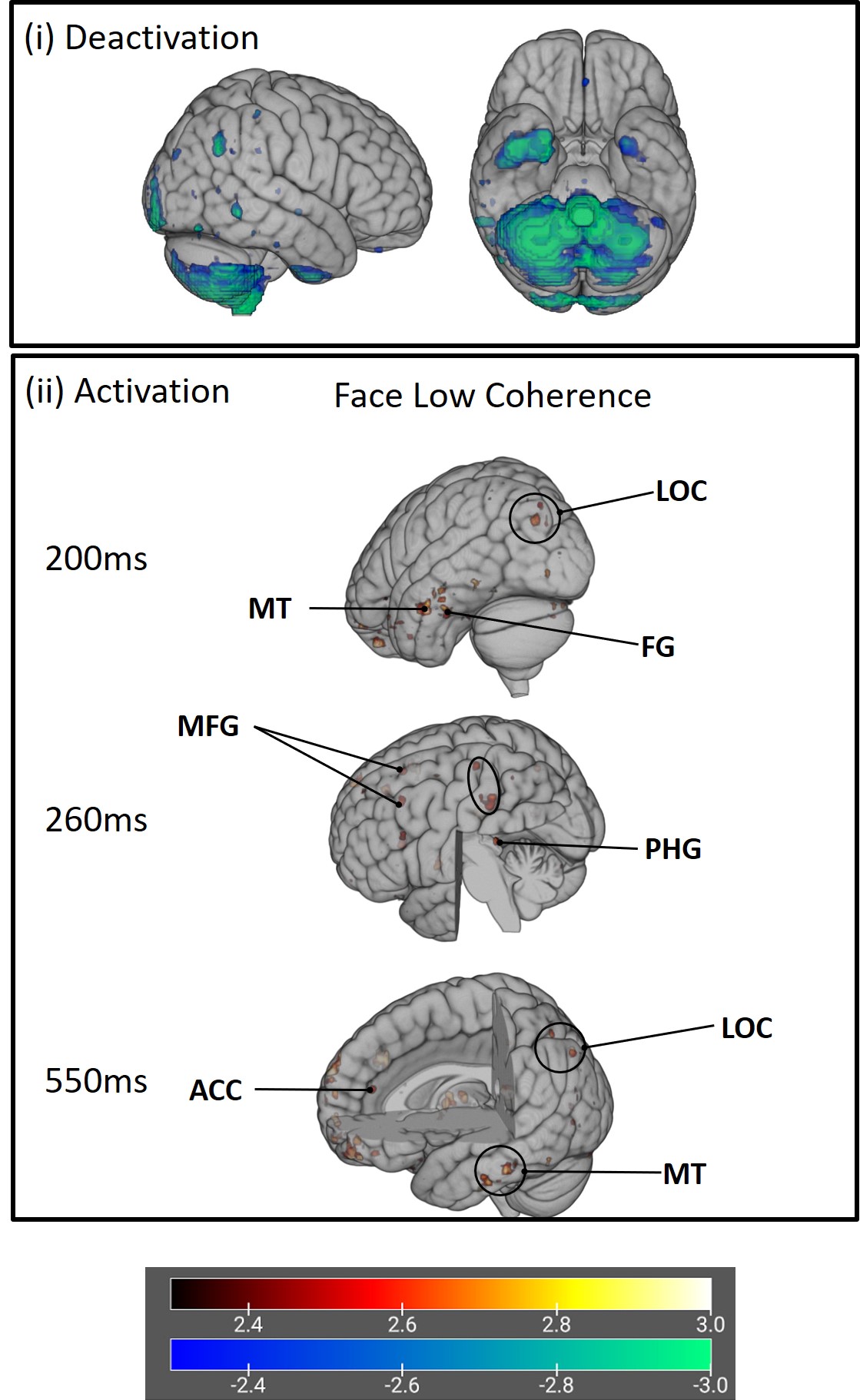}
  \caption{Representative time frames of  deactivation (i)  and activation (ii) in the spatio-temporal latent source space  (uncorrected Z-maps) of Three-choice Visual Categorization Task under face category low-coherence setting.}
  \label{FCHCompactResult}
\end{figure}  

\subsubsection{Face-Car-House visual categorization task}
 
Similarly, for the visual categorization task, we also reconstructed a high-resolution latent source space with a spatiotemporal resolution of 2 mm $\times$ 2 mm $\times$ 2 mm $\times$ 100 Hz. In particular, we present representative time frames of the source space for the low-coherence face trials in Fig. \ref{FCHCompactResult} where faces are more salient stimuli and low level of sensory evidence increases the decision ambiguity. Similar activation patterns were also observed in other task conditions. As shown in Fig. \ref{FCHCompactResult}ii, early activations around 200 ms post-stimulus occur in the middle temporal gyrus (MT), lateral occipital cortex (LOC), and fusiform gyrus (FG). These regions have been implicated as the neural constituents of sensory processing and decision formation, in line with findings in previous studies on the same dataset analyzed using asymmetrical fusion methods \cite{muraskin2018multimodal, tu2017network}. At 260 ms post-stimulus, in addition to the activation in the middle frontal gyrus (MFG) which was also reported in \cite{muraskin2018multimodal} as part of the decision monitoring system, our method revealed new activations in the sensory and motor cortices related to planning and executing of button pressing and the parahippocampal gyrus (PHG). This area was likely activated due to the additional memory processing required when sensory evidence is ambiguous. At around 550 ms, we observed activations in the anterior cingulate cortex (ACC), lateral occipital cortex (LOC), and middle temporal gyrus (MT), consistent with the reactivation of neural network hypothesis implicated by \cite{muraskin2018multimodal}.

\subsection{Recovering sources in simulated data}\label{subsec3}

We designed a realistic simultaneous EEG-fMRI simulator to validate our method under different acquisition parameters and noise levels. Since the cyclic transcoder is able to estimate an EEG latent source space and an fMRI latent source space from the data separately, the performance of our method on recovering the latent source space and the EEG-fMRI observations estimated from each other was evaluated from both EEG-to-fMRI and fMRI-to-EEG transcoding pathways.

\begin{figure}[thpb]
  \centering
  \includegraphics[scale=0.35]{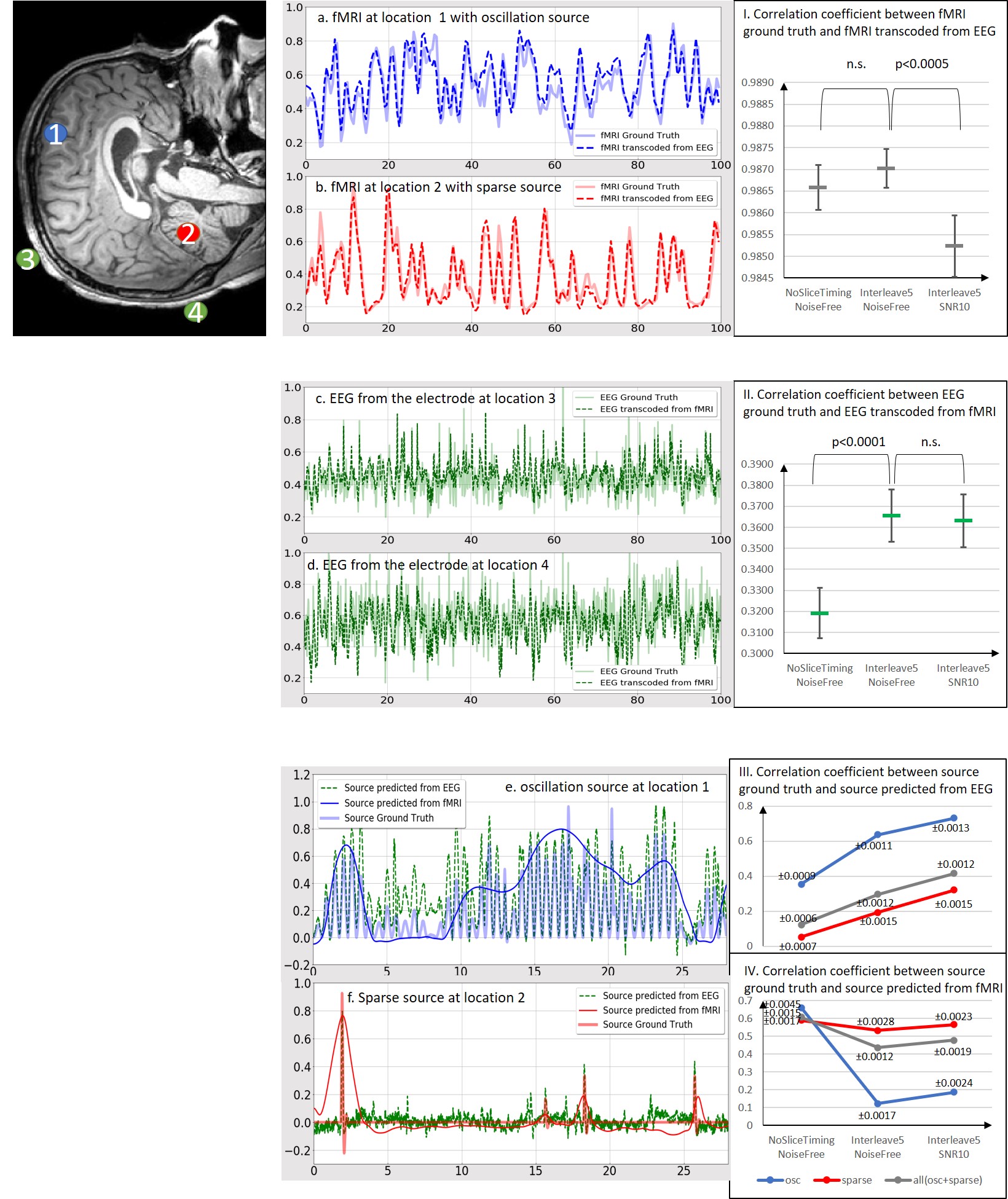}
  \caption{Result showing transcoder predictions relative to ground truth: The models are trained with data of different slicetiming settings and noise levels. Example neural signals from location 1  (oscillation source) , location 2 (sparse source)  and two electrodes 3 and 4 on the surface of the scalp are shown in subplots a to f. Subplot I: the correlation coefficient between the fMRI estimation of EEG-to-fMRI transcoder and ground truth fMRI,95\% confidence intervals are shown as error bars. Example fMRI signals are shown in Subplot a and b. Subplot II: the correlation coefficient between the source estimation of EEG-to-fMRI transcoder and ground truth source. $\pm$ indicates 95\% confidence intervals at each point. Example source signals are shown in subplots c and d. Subplot III: the correlation coefficient between the source estimation of fMRI-to-EEG transcoder and ground truth source. $\pm$ indicates 95\% confidence intervals at each point. Example source signals are shown in Subplot c and d. Subplot IV: the correlation coefficient between the EEG estimation of fMRI-to-EEG transcoder and ground-truth EEG, 95\% confidence intervals are shown as error bars. Example EEG signals are shown in Subplot e and f.}
  \label{transcoder predictions}
\end{figure}

Fig.\ref{transcoder predictions}i-iv shows the mean values of correlation coefficient between the model predictions and ground-truth signals averaged across 19 runs of test data with a duration of 600 s under three different scenarios.
Specifically, our method was able to reconstruct the fMRI observations via the EEG-to-fMRI transcoder with high fidelity across all scenarios (Fig.\ref{transcoder predictions}i). Similarly, the reconstructed EEG observations from the fMRI-to-EEG transcoder (shown in Fig.\ref{transcoder predictions}ii) also showed significant correlations with the ground-truth. Fig.\ref{transcoder predictions}a and Fig.\ref{transcoder predictions}b show the reconstructed EEG and fMRI observations at representative electrodes and locations overlaid with the corresponding ground-truth signals. To evaluate the effect of noise and interleave slice timing, we also simulated data with 5 interleave slice-timing and SNR10. For both reconstructions, results in the interleave 5 noise-free scenario produced the highest correlation probably because the interleaved slice acquisition gives the model access to more variability in the training data. When noise was added, the performance degraded (interleave5 noise-free vs. SNR10).

Fig.\ref{transcoder predictions}iii shows the mean correlation coefficients between the EEG latent source estimates and the ground-truth, calculated from averaging 304 epochs of held-out data with a duration of 30 s. The mean correlation coefficients are shown separately for the sources of oscillatory and sparse activity. Similar results of the fMRI source estimates are shown in Fig.\ref{transcoder predictions}iv. Fig.\ref{transcoder predictions} c-f show the reconstructed EEG and fMRI source activity at representative locations.
While adding the interleave 5 slicetiming improved the performance of EEG latent source reconstruction in the EEG-to-fMRI transcoder, it led to reduced accuracy in fMRI latent source reconstruction in the fMRI-to-EEG transcoder. This is likely because inferring the source from fMRI requires upsampling and deconvolution in the temporal dimension, both of which are sensitive to timing. The interleaved acquisition design adds uncertainty in the temporal dimension. Inferring the source from EEG, on the other hand, only involves spatial deconvolution, where timing is not critical.

\begin{figure}[thpb]

  \centering
  \includegraphics[scale=0.25]{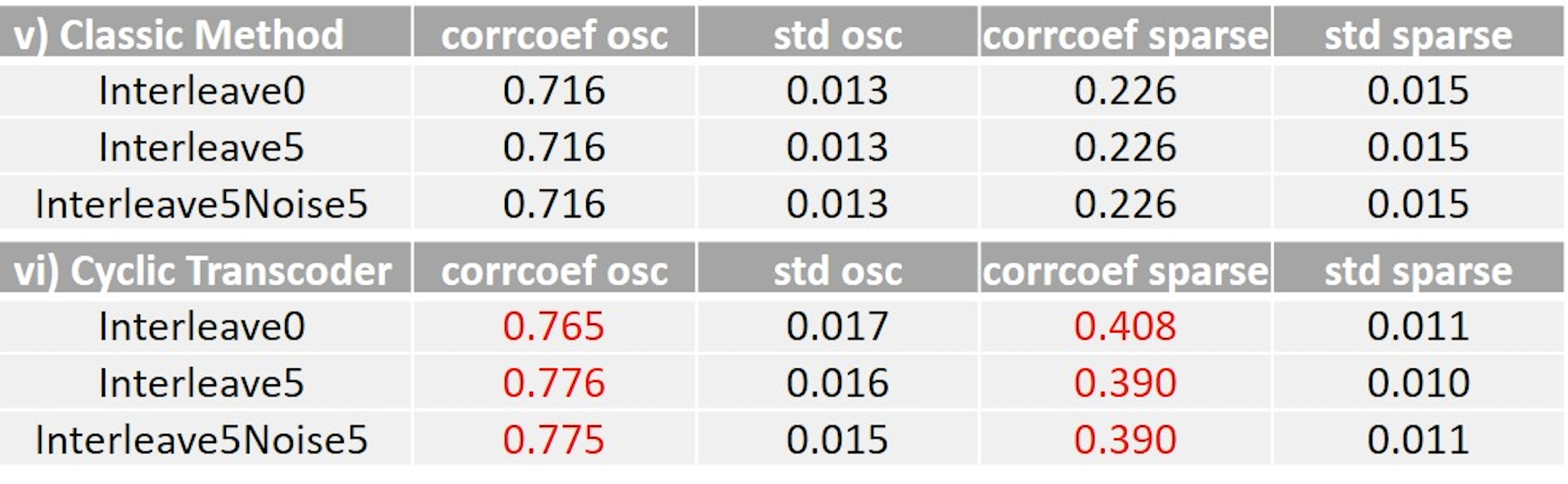}
  \caption{Compare the performance of Cycle-CNN transcoder and vanilla transcoder: i) EEG source estimation correlation coefficient and standard deviation of Classic EEG sourcing method. ii) EEG source estimation correlation coefficient and standard deviation of Cycle-CNN transcoder.}
  \label{SimulationResults}
\end{figure}

Additionally, we compared the performance of our method with the classic minimum-norm EEG source localization (MNE) method \cite{molins2008quantification} on the reconstruction of latent EEG sources. In particular, to favor the classic method, we provided MNE method the ground-truth leadfield matrix which typically has to be estimated from the data in practice. Fig.\ref{SimulationResults} shows the comparison of Pearson correlation coefficients between the estimated EEG sources and simulated sparse and oscillatory EEG sources for both methods. We demonstrated in Fig.\ref{SimulationResults} that our method outperforms the classic MNE method even when MNE has access to the ground-truth leadfield matrix.

Taken together, our model is capable of resolving the latent source space and reconstruct EEG-fMRI observations with high fidelity. Simulation results demonstrate our model's capability of capturing the shared underlying neural substrate between EEG and fMRI. Such capability enables the model to map from one modality to the other while leveraging the complementary spatial and temporal information in two modalities to achieve enhanced resolution in both domains.

\section{Discussion}\label{sec12}
We have developed a new framework for symmetric fusion of simultaneously acquired EEG-fMRI data - a hierarchical deep transcoder, with a decoder-encoder architecture using convolutional layers, to enable the transcoding from one modality to the other via a highly spatiotemporal-resolved latent source space. Our approach is completely data-driven and does not use prior information on the biophysical generative process of each modality. As a result, 
it is able to capture nonlinear relationships between EEG and fMRI, in contrast to many previous methods which hinge on linearity assumptions in the cross-modal integration. Once the model training is complete, we can obtain estimates of latent source space decoded from EEG or fMRI, HRF, EEG forward and inverse models from the two transcoding pathways at once. We demonstrate the superior performance of our model on the source reconstruction problem using extensive realistic simulations where the ground-truth source space is known. Specifically, our model achieves more accurate estimates of the underlying source space, outperforming the conventional EEG source localization method. Using a hierarchical transcoding framework, our model is capable of reconstructing a group-level and a subject-level source spaces with different resolutions while accommodating the inter-subject variability. The ability of the model in recovering full-brain source spatiotemporal dynamics on real EEG-fMRI data is a major advancement of our approach. We demonstrate that our model can not only reproduce previous findings on the source dynamics underlying two event-related cognitive tasks but also uncover new source dynamics that are otherwise impossible to achieve with previous methods. Moreover, the trained model can be applied to improve source localization in scenarios where only single modality data is available, making our model useful in widespread applications such as brain-computer interface (BCI) or fMRI connectivity analysis. 

Asymmetric fusion methods EEG-informed fMRI or fMRI-informed EEG analyses rely on the covarying activity between EEG and fMRI latent sources and use information from each modality unequally. The activity in one modality which is not shared by the other modality will either be neglected or incorporated to produce biased fusion results. Our model, on the contrary, is a generative approach and relies on the shared information between both modalities only in the training stage. Once the training is completed, the latent source spaces, estimated from EEG and fMRI, become independent of each other, and thus are capable of preserving modality-specific information that might contribute to an optimized cross-modal fusion. In other words, upon fusing these two modalities, the integrated latent source space captures both shared and modality-specific information. It is therefore highly likely that our approach is capable of identifying new spatiotemporal brain dynamics with less bias towards either modality. For example, the dissociation of early and late brain networks from the EEG-informed fMRI analysis in \cite{muraskin2018multimodal} hypothesized that such dissociation related to the EEG contrast between the level of sensory evidence (high vs. low coherence levels). Without explicit hypothesis about the task, our approach can uncover broader findings in the visual categorization task, e.g., prolong deactivation, activation at motor and sensory cortex and parahippocampal gyrus. The later is also an indication that our method can access information outside of the common neural substrate of EEG and fMRI as it is broadly believed to be problematic extracting information from regions deep in the brain (e.g. parahippocampal gyrus) by EEG \cite{grech2008review} ("I am not sure if this claim is accurate"). It is also noteworthy that activation maps shown in Section \ref{real data results} have far less statistical power for every time frame compared with a typical fMRI’s activation map generated from the general linear model (GLM) analysis. Specifically, our method computes an activation map every 10 ms. With a typical temporal resolution of fMRI at 2 s, the activation map in GLM can be considered as blocks of 2 s data collapsed together and thus has a statistical power about 200 times of our method. Because of that, the high-resolution latent source space may fail to show some of the activation/deactivation that shows in an fMRI activation map under the same threshold with the same number of trials. On the other hand, it may be able to capture some of the activation/deactivation which could be cancelled out with each other in a typical fMRI analysis without this collapsing effect.

Our framework is not merely a "black-box" approach in that we dissociate spatial and temporal transformations on EEG and fMRI by incorporating the biophysical knowledge about their own generative processes. The fMRI decoder, which applies only temporal transpose convolution, solves the fMRI deconvolution problem. The fMRI encoder, on the other hand, consisting of only temporal convolutional layers, learns the HRF functions for mapping the latent source to the fMRI measurement space from the training data. Likewise, the EEG encoder with only spatial convolutional layers and the EEG decoder using only transpose spatial convolutional layers implement an EEG forward model and an EEG inverse model, respectively. Such dissociation makes the transcoder robust to noise only present in one modality. For example, the cardiac noise in fMRI is notoriously difficult to remove due to its large overlap in frequency range with the actual signal. Since only the information shared by EEG and fMRI drives the model training process, as long as EEG data were cleaned without cardiac related noise, the fMRI decoder applying only temporal transformations will learn temporal filters resistant to the cardiac noise from EEG estimated source while preserving the high spatial specificity of fMRI. Similarly, the EEG decoder, which only applies spatial transformation, is able to learn spatial filters that can remove electrooculography (EOG) artefacts since this type of artefacts is not present in fMRI estimated source.

Ideally, one would directly train a group-level source model with the desired high spatiotemporal resolution. In practice, however, we choose to train a group-level at an intermediate spatial resolution of 12 mm $\times$ 12 mm $\times$ 12 mm and temporal resolution of 2.7 Hz due to the constraints of computing power and memory capacity. To this end, EEG data need to be downsampled temporally from 500 Hz to 2.7 Hz, and fMRI data need to be dowsnampled spatially to 12 mm $\times$ 12 mm $\times$ 12 mm from 2 mm $\times$ 2 mm $\times$ 2 mm before feeding into the transcoder. Although the downsampling process causes information loss in one of the two dimensions for EEG and fMRI, it is necessary to make the training of the group-level model feasible, which subsequently leads to gain in improved resolution in the other dimension for both EEG and fMRI, i.e., the spatial resolution of EEG increases to 12 mm $\times$ 12 mm $\times$ 12 mm and the temporal resolution of fMRI increases to 2.7 Hz. It is arguably a reasonable choice since fMRI is not able to capture high frequency brain activity (higher than 5 Hz). Once the model is trained, this gain in temporal resolution for EEG and spatial resolution for fMRI will be preserved during inference where the downsampling preprocessing is not required. It is noteworthy that the choice of the intermediate resolution for the group-level source space is a tunable hyperparameter which largely depends on the available computational resources. The desired 2mm/100Hz resolution is achieved in training of a subject-level model where the group-level EEG/fMRI source estimates are epoched and averaged across trials to increase SNR. The averaging is particularly useful for fMRI since it can leverage the jittering event-related design to increase the information density in the sparsely sampled fMRI activity before it is upsampled to the desired high temporal resolution \cite{amaro2006study}. The epoched group-level EEG and fMRI source estimates are also sliced along the spatial and temporal dimensions into small trunks to make training the subject-level model feasible given the computation resource constraints. Similarly, the slicing operation is not necessary during inference and the trained subject-level model can be applied to each time point/3D volume for EEG and/or fMRI in original temporal/spatial resolution.

Since the ground-truth brain source activity is not accessible, another major contribution of our work is the development of a realistic EEG-fMRI simulator to generate synthetic datasets, which are useful for rigorous assessments of the robustness and accuracy of the transcoder model in recovering the latent source space from EEG and fMRI observations. These simulations were designed to capture various characteristics of the real modality-specific data via the generative modeling of EEG and fMRI, such as various of commonly observed artefacts (slice-timing effect, physiological noise, and motion artefacts) in fMRI. In particular, the simulator enables the simulations of two types of source activity (oscillatory vs. event-related sparse), which can also be designed to vary across different brain regions to simulate the regional variability. While many studies have employed simulated EEG or fMRI for algorithm validation, no prior work has developed a flexible and comprehensive simultaneous EEG and fMRI simulator to enable the evaluation of modality fusion methods in a controlled manner using ground truth as well as varying experimental setup. One limitation of the simulator, however, lies in that it can not simulate structured cross-regional brain dynamics and subject-wise difference. It is therefore not suitable to evaluate the subject-level transcoder. Future work will incorporate simulations of task-related brain dynamics in the absence of stimuli and subject-wise variability to enable a more rigorous validation of our model on both levels.

Moreover, the trained model can be applied to EEG or fMRI data that are not collected in the same session. If only EEG is acquired in one experimental session, one can apply the trained model to generate the latent source space estimated from EEG, and use such source space to transcode EEG data to the space of fMRI signal. Similarly, if only the fMRI data are available, one can transcode them into EEG via the latent source space estimates from fMRI. For instance, without any simultaneous EEG-fMRI recording of a subject, using the trained group-level transcoder only, we can resolve EEG’s temporal resolution for any subject to around 12 mm. If a subject participates in a 60-min simultaneous EEG-fMRI recording, we can train the subject-level model for the subject which can potentially boost EEG source localization resolution to 2mm. More importantly, this subject-level model can be applied to the same subject’s other EEG data and to solve for EEG source localization with high accuracy. Such improved EEG source localization in resolution can be very beneficial to real-time BCI applications. It is also possible that our model can be used as a low cost, computationally-driven approach to produce fMRI images from EEG recordings, thus enabling a \$600 - \$1200 scan to be done at a cost of $<$ \$10.

\section{Methods}\label{sec11}

\subsection{Simlatenous EEG-fMRI datasets}
We evaluated the performance of the transcoding model on two simultaneous EEG-fMRI datasets recorded when subjects performed an auditory oddball task and an event-related three-choice visual categorization task.
\subsubsection{Auditory oddball task}
19 healthy human subjects performed an auditory oddball task \cite{hong2020mapping}, with a 8:2 ratio of standard versus oddball  (target)  stimuli. Standard stimuli were pure tones with a frequency of 350 Hz, while the oddball stimuli were broadband  (laser gun)  sounds. Each stimulus has a duration of 200 ms, with an inter-trial interval  (ITI) sampled from a uniform distribution between 2 s and 3 s. Subjects were instructed to attend only to oddball sounds and respond with a button press as quickly and as accurately as possible.
%(Figure \ref{AuditoryOddballTask} in Supplementary). 
By design, each subject was to complete five sessions with 105 trials per session. The actual sessions completed per subject was 4.6 sessions  (with a range of 2 to 5, and a standard deviation of 0.98).

EEG was recorded using a 64-channel BrainAmp MR Plus system (Brain Products). MR data were recorded inside a 3 T Siemens Prisma scanner with a 64-channel head/neck coil. Specifically, structural T1 images were acquired with an echo time  (TE)  of 3.95 ms, a repetition time  (TR)  of 2300 ms, a 1 mm in-plane resolution and 1 mm slice thickness. Functional Echo Planar Imaging  (EPI)  images were acquired with a TE of 25 m, a TR of 2100 ms, a 3 mm in-place resolution and 3 mm slice thickness.

\subsubsection{Face-Car-House visual categorization task}

21 healthy subjects performed a event-related three-choice visual categorization task. On each trial, an image of a face, car, or house was presented for 50 ms. Subjects reported their choice of the image category by pressing one of the three buttons on an MR-compatible button response pad with three fingers on their right hand. The stimuli consisted of a set of 30 face, 30 car, and 30 house images. The phase coherence of the images was degraded at a high coherence (50\%) level and at a low coherence (35\%) level by a weighted mean phase algorithm. The stimuli display was controlled by E-Prime software (Psychology Software Tools) using a VisuaStim Digital System (Resonance Technology) with a $600 \times 800$ pixel goggle display. Images subtended $11 \times 8^{\circ}$ of visual angle. Each subject participated in four runs of the categorization task. In each run, there were 180 trials (30 per condition; 6 conditions: face high, car high, house high, face low, car low, and house low). The inter-trial interval was sampled uniformly between 2 and 4 s. Therefore, data from 720 trials (240 of each category and 360 of each coherence) were acquired for each subject during the entire experiment.

EEG data were recorded simultaneously with the fMRI data using a custom-built MR-compatible EEG system \cite{goldman2009single, sajda2010signal} with differential amplifiers and bipolar EEG montage, using a 1 kHz sampling rate. The caps were configured with 36 Ag/AgCl electrodes, including left and right mastoids, arranged as 43 bipolar pairs. Further details of the recording hardware are described by Sajda et al. \cite{sajda2010signal}. Functional echo-planar image data were collected using a 3T Philips Achieva MRI scanner (Philips Medical Systems) with the following scanning parameters: TR =  2000 ms; TE =  25 ms; flip angle, 90°; slice thickness, 3 mm; interslice gap, 1 mm; in-plane resolution, $3 \times 3$ mm; 27 slices of $64 \times 64$ voxels per volume; 280 total volumes. For all participants, a high-resolution structural image was also acquired using spoiled gradient recalled echo sequence with a $1 \times 1 \times  1$ mm resolution and 150 slices of $256 \times 256$ voxels.

\subsubsection{Data preprocessing}

Raw EEG data were preprocessed off-line using Matlab (Mathworks) following a well-adopted standard pipeline. Band-pass filter and notch filters were applied in a non-causal zero-phase form to remove direct current drift, electrical line noise, and high-frequency artifacts not associated with neurophysiological processes. Gradient artifacts were removed using an fMRI Artifact Slice Template Removal algorithm (FASTR) \cite{niazy1999improved}. To remove ballistocardiogram (BCG) artifacts, we adopted a conservative approach, based on principal component analysis, where a small number of principal components that captured BCG artifacts were selected for each subject from the gradient-free EEG data. These components were then projected back on to the broadband data and subtracted out to produce the BCG-free data.

fMRI data were preprocessed using FSL (www.fmrib.ox.ac.uk/fsl/). The preprocessing steps include slice-timing correction, motion correction, spatial smoothing, and high-pass filtering ($>$100 s). Functional images were first transformed into each subject’s high-resolution anatomical space using boundary based registration \cite{greve2009accurate}, and then spatially normalized to the standard Montreal Neurological Institute brain template using FAST (Oxford Centre for Functional MRIs’ Automated Segmentation Tool \cite{zhang2001segmentation}). Details of the preprocessing steps for each dataset are described in Supplementary Material.

\subsection{Simultaneous EEG-fMRI simulator} \label{simulator}
Due to the lack of ground-truth latent source space activity in real simultaneous EEG-fMRI data, we developed a realistic simultaneous EEG-fMRI simulator to generate simulations, which were used for evaluating the performance of our algorithm in the source reconstruction task from EEG and fMRI.

Specifically, we assume that latent neural sources $\mathbf{X}_\text{t}$ are uniformly distributed in a 3D volumetric space in the brain. The latent source at a specific spatial location was modeled as a train of impulse functions, corresponding to the transient evoked brain responses. Source activity $\mathbf{X}_\text{t}$ was then transformed into EEG and fMRI observations via their own linear forward model: 

\begin{equation}\label{eq1}
\mathbf{F}_\text{t}=\mathbf{X}_\text{t}\circledast \mathbf{h}+\mathbf{N}_\text{F}
\end{equation}
\begin{equation}\label{eq2}
\mathbf{P}_\text{t}=\mathbf{X}_\text{t}\circledast \mathbf{d}
\end{equation}
\begin{equation}\label{eq3}
\mathbf{E}_\text{t}=\mathbf{G}\mathbf{P}_\text{t} + \mathbf{N}_\text{E}
\end{equation}
In our simulation, source activity $\mathbf{X}_\text{t}$ is a $64\times 64 \times 35 \times \text{t}$ matrix sampled at a frequency of 105 Hz. The size of $\mathbf{X}$ was chosen according to the typical size of a real ($\text{T2}^*$) BOLD image and the sampling rate was chose to cover most of EEG frequency bands of interest. $\mathbf{F}_\text{t}$ is the hemodynamic response (BOLD) modeled as a convolution of the latent source activity $\mathbf{X}_\text{t}$ with canonical hemodynamic impulse response functions (HRFs) $\mathbf{h}$. Therefore, $\mathbf{F}_\text{t}$ is also a 4D matrix with the same spatial dimension as $\mathbf{X}_\text{t}$ but it is downsampled at every 2 s for each voxel, as in a typical EEG-fMRI experiment. $\mathbf{N}_\text{F}$ represents various sources of noise present in real fMRI signals including the respiratory noise, cardiac noise, and background Gaussian noise. The EEG event-related potential $\mathbf{P}_\text{t}$ is modeled as a convolution between the source activity $\mathbf{X}_\text{t}$ and the "potential impulse response function" $\mathbf{d}$. In contrast to the sluggish nature of $\mathbf{h}$, $\mathbf{d}$ is a much briefer impulse response with a duration around 470 ms. The scalp EEG observation $\mathbf{E}_\text{t}$ is a $64 \times \text{t}$ 2D matrix which represents a linear mixing of $\mathbf{P}_\text{t}$ through a leadfield matrix $\mathbf{G}$ with additive Gaussian noise $\mathbf{N}_\text{E}$. It has the same temporal resolution as $\mathbf{X}_\text{t}$, but a much lower spatial resolution. $\mathbf{G}$ can be estimated for each subject from their anatomical $\text{T1}$ image using  the  Boundary  Element  Method  (BEM) implemented in the FieldTrip toolbox \cite{oostenveld2011fieldtrip}. To make our simulations more realistic, we incorporated regional variations in latent sources based on their anatomical locations, which can be determined by registering the subject's structural $\text{T1}$ image to their corresponding functional image ($\text{T2}^*$) space. In particular, we added oscillatory activity on top of those evoked responses, across various frequency ranges, to different areas of the brain. We also varied the magnitude of noises at different spatial locations, added motion artifacts, and simulated slice-timing effects in fMRI signals. A total 319 runs of 10 min (500 epochs training chucks of 30, 300 TRs) simultaneous EEG-fMRI simulated datasets were generated with this simulator and were divided into a training set of 300 runs and a test set of 19 runs. More details of the simulator are described in Supplementary Material.

\subsection{Multi-scale deep transcoder for modality fusing}
\subsubsection{Algorithm overview}
Our goal is to solve for the latent source activity  $\mathbf{X}_\text{t}$ from simultaneous EEG-fMRI observations $\mathbf{E}_\text{t}$ and $\mathbf{F}_\text{t}$. This source reconstruction problem can be expressed as:
\begin{equation}\label{optimization}
{\mathbf{\hat X}}_\text{t} = \arg\min_{\mathbf{X}_\text{t} }\frac{\alpha}{2}\left \| \mathbf{E}_\text{t}-\mathbf{G}\mathbf{P}_\text{t}  \right \|+
\frac{1-\alpha}{2}\left \| \mathbf{F}_\text{t}-\mathbf{X}_\text{t}\mathbf{H}  \right \| + \lambda \phi(\mathbf{X}_\text{t})
\end{equation}
where $\alpha$ tunes the trade-off between the two modalities and $\lambda$ controls the regularization term $\phi(\mathbf{X}_\text{t})$. Without access to accurate a priori estimates of the HRFs $\mathbf{H}$ and leadfield matrix $\mathbf{G}$, reconstructing $\mathbf{X}_\text{t}$ from EEG $\mathbf{E}_\text{t}$ can be regarded as a blind signal separation (BBS) problem and reconstructing $\mathbf{X}_\text{t}$ from fMRI $\mathbf{F}_\text{t}$ can be viewed as a blind deconvolution problem. To symmetrically fuse EEG and fMRI data while reconciling the difference in spatial/temporal resolution between the two modalities, we used a multi-scale deep cyclic convolutional transcoder model, which comprises an fMRI-to-EEG decoder-encoder and an EEG-to-fMRI decoder-encoder. This model is capable of solving the source reconstruction problem via spatial and temporal convolutional operations without any prior knowledge of EEG and fMRI forward models. Since we don't have access to the ground-truth latent source activity during the training phase, the model learning is therefore self-supervised and the task is to minimize the reconstruction loss of EEG and fMRI signals concurrently. The multi-scale transcoding process consists of two stages. In the first stage, we train a group-level model to map EEG and fMRI into a source space with an intermediate spatial and temporal resolution. This step is necessary because the cost for computation and storage would otherwise become prohibitive if we were to directly map into the source space with desired spatiotemporal resolution (2 mm/100 Hz). In the second stage, we train a subject-level model for each subject individually, to reach the desired spatiotemporal resolution. Such reconstructed sources are considered in the super-resolution latent source space after accounting for subject-wise variability.

\subsubsection{Cyclic convolutional transcoder}
The core component of our multi-scale deep transcoding model is a cyclic convolutional transcoder. Both of the two stages use the same cyclic convolutional transcoder model architecture consisting of four modules: fMRI decoder, EEG encoder, EEG decoder, and fMRI encoder as shown in Fig. \ref{Method}. The fMRI decoder and EEG encoder form the top fMRI-to-EEG transcoding pathway where fMRI data is first temporally upsampled to arrive at a fMRI-estimated source space. The fMRI-estimated source is then translated into EEG through EEG encoder by applying spatial convolution. Similarly, at the bottom EEG-to-fMRI transcoding pathway, EEG is first passed to the EEG decoder to get an EEG-estimated source via spatial upsampling, from which fMRI is then generated through the fMRI encoder by temporal convolution. Upsampling in space/time is achieved using spatial /temporal transposed convolution layer. Note that in EEG (fMRI) decoder the transposed convolution is only applied in time (space), the same principle holds for the fMRI (EEG) encoder, where the convolution is only applied in space (time). This spatial and temporal separation ensures the interpretability of our model: the EEG decoder learns an EEG inverse model, the fMRI decoder solves for fMRI deconvolution, the EEG encoder estimates the leadfield matrix in EEG, and the fMRI encoder models the hemodynamic response function. The EEG/fMRI encoder (decoder) is composed of 5 stacked convolutional layers. Each convolutional layer consists of spatial (temporal) convolution or transposed convolution with rectified linear unit (ReLU) as activation function, followed by batch normalization (BN) layer in the group-level model and residual layer in the subject-level model. 

In addition to the two parallel transcoding pathways (fMRI-to-EEG and EEG-to-fMRI), we also employ two cross encoding pathways (EEG-to-EEG and fMRI-to-fMRI) to encourage cycle consistency \cite{zhu2017unpaired}. These two additional pathways further regularize the self-supervision process resulting in the total loss as a sum of four reconstruction losses:
\begin{equation}\label{loss}
\text{L} = \text{L}_{\text{fMRI} \rightarrow \text{EEG}} + \text{L}_{\text{EEG} \rightarrow \text{fMRI}} + \text{L}_{\text{fMRI} \rightarrow \text{fMRI}} + \text{L}_{\text{EEG} \rightarrow \text{EEG}} 
\end{equation}

\begin{itemize}
\item fMRI-to-EEG transcoding loss: $\text{L}_{\text{fMRI} \rightarrow \text{EEG}} = \sum_{i=1}^{n}(\mathbf{E}_i-\hat{\mathbf{E}_i})^2$ 

\item EEG-to-fMRI transcoding loss: $\text{L}_{\text{EEG} \rightarrow \text{fMRI}} = \sum_{i=1}^{n}(\mathbf{F}_i-\mathbf{\hat{F}}_i)^2$

\item fMRI-to-fMRI cycle consistency loss: $\text{L}_{\text{fMRI} \rightarrow \text{fMRI}} = \sum_{i=1}^{n}(\mathbf{F}_i-\mathbf{\hat{F}'}_i)^2$

\item EEG-to-EEG cycle consistency loss: $\text{L}_{\text{EEG} \rightarrow \text{EEG}} = \sum_{i=1}^{n}(\mathbf{E}_i-\mathbf{\hat{E}'}_i)^2$

\end{itemize}
where $\mathbf{E}_\text{i}$ and $\mathbf{F}_\text{i}$ are the input EEG and fMRI, $\mathbf{\hat{E}}_i$ is the transcoded EEG from fMRI, $\mathbf{\hat{F}}_i$ is the transcoded fMRI from EEG, $\mathbf{\hat{E}'}_i$ is the reconstructed EEG from EEG-to-EEG cycle, and $\mathbf{\hat{F}'}_i$ is the reconstructed fMRI from fMRI-to-fMRI cycle.

Although the group-level model and the subject-level model use the same model architecture, their model parameters are not shared during training. As a result, the model parameters in the subject-level model need to be re-initialized for each subject. 

\subsubsection{Two-stage source reconstruction}
To recover latent sources with the desired spatiotemporal resolution (2 mm/100 Hz), the reconstruction process entails upsampling raw EEG data in space approximately by a factor of 2,240 times and raw fMRI data in time by a factor of 200 times. In practice, the accompanying heavy computation cost makes it infeasible to combine data across all subjects during model training. Consequently, we designed a two-stage training pipeline to first train a group-level transcoder model to arrive at an intermediate spatiotemporal resolution by combing all subjects, and then fine-tune a subject-level transcoder model at the desired resolution for each individual subject. This hierarchical approach enables training the model with more data to reduce over-fitting, as well as fine-tuning the model to take into account subject-wise variability. 

Prior to training the group-level model, artifacts-free simultaneous EEG-fMRI data were first spatially or temporally interpolated to an intermediate spatial/temporal resolution. The interpolated EEG/fMRI data were then mean-averaged in time/space so that they were sampled at the same intermediate spatial/temporal resolution of approximately 12 mm/2.7 Hz as the source space (varies across subjects based on the fMRI slice-timing setting). Although we compromised the spatial resolution of fMRI and the temporal resolution of EEG via the mean-averaging operation to achieve a 'middle-ground' resolution in order to train the group-level model, it is worth noting that this compromise in resolution is only required during train. During inference, the model takes as input the linearly interpolated EEG/fMRI and generates a group-level EEG-estimated source with a spatiotemporal resolution of 12 mm/100 Hz and a group-level fMRI-estimated source with a spatiotemporal resolution of 2 mm/2.7 Hz separately.

In training of the subject-level model, our goal is to map the group-level reconstructed sources to the desired super-resolution for each subject. To this end, we first performed linear interpolation on the group-level EEG/fMRI-estimated sources in space/time to achieve a spatial/temporal resolution of 2 mm/ 100 Hz. As opposed to feeding the entire continuous data volume to the transcoder as in the group-level training, we extracted epochs time-locked at the stimulus onset with a duration of 3 s (0 s prestimulus to 3 s poststimulus) from the group-level EEG-estimated and fMRI-estimated sources. The group-level EEG-estimated and fMRI estimated source epochs were then averaged across trials to produce two single 4D tensors with a dimension of $64\times 64 \times 35 \times 300$. This grand-averaging step is crucial to increase the 
SNR for EEG/fMRI and alleviate the computational burden. To satisfy the GPU memory constraint, these two 4D tensors were decomposed into overlapping chucks in space and in time $16 \times 16 \times 16 \time 200$ before feeding into the subject-level model. For each subject, we trained a subject-level model separately. During inference, similar operations were applied to obtain an EEG-estimated source and a fMRI-estimated source,  both of which are resolved at the desired spatiotemporal resolution.

Once the group-level and the subject-level transcoder models are trained and combined, for each subject, we can apply them directly to continuous (un-epoched) simultaneous EEG-fMRI data to reconstruct a EEG-estimated source and a fMRI-estimated source, both of which are also continuous in time with the same spatiotemporal resolution of 2 mm/100 Hz and can be added up together to form a single super-resolution latent source space.

\subsection{Training and evaluation}

\subsubsection{Simulations}
We simulated 319 runs of 600 second simultaneous EEG-fMRI data. Data are cut into 30 s chunks with 50\% overlap. 300 of 319 runs are used for training the model, with the remaining 19 runs used for testing the model. The simulated data is used in evaluating the transcoding algorithm he network is trained with batchsize set to 1 by backpropagation and the gradient-based optimisation is performed using the Adam optimizer for 32 epoches.

% \bibliographystyle{spphys}       % APS-like style for physics
% \bibliography{sn-bibliography} 
%\begin{thebibliography}{}
% etc
%\end{thebibliography}

\section{Supplementary material}

under construction

\textbf{Ethical approval declarations} (only required where applicable) Any article reporting experiment/s carried out on (i)~live vertebrate (or higher invertebrates), (ii)~humans or (iii)~human samples must include an unambiguous statement within the methods section that meets the following requirements: 

\begin{enumerate}[1.]
\item Approval: a statement which confirms that all experimental protocols were approved by a named institutional and/or licensing committee. Please identify the approving body in the methods section

\item Accordance: a statement explicitly saying that the methods were carried out in accordance with the relevant guidelines and regulations

\item Informed consent (for experiments involving humans or human tissue samples): include a statement confirming that informed consent was obtained from all participants and/or their legal guardian/s
\end{enumerate}

If your manuscript includes potentially identifying patient/participant information, or if it describes human transplantation research, or if it reports results of a clinical trial then  additional information will be required. Please visit (\url{https://www.nature.com/nature-research/editorial-policies}) for Nature Portfolio journals, (\url{https://www.springer.com/gp/authors-editors/journal-author/journal-author-helpdesk/publishing-ethics/14214}) for Springer Nature journals, or (\url{https://www.biomedcentral.com/getpublished/editorial-policies\#ethics+and+consent}) for BMC.
\backmatter

\bmhead{Supplementary information}

If your article has accompanying supplementary file/s please state so here. 

Authors reporting data from electrophoretic gels and blots should supply the full unprocessed scans for key as part of their Supplementary information. This may be requested by the editorial team/s if it is missing.

Please refer to Journal-level guidance for any specific requirements.

\bmhead{Acknowledgments}

Acknowledgments are not compulsory. Where included they should be brief. Grant or contribution numbers may be acknowledged.

Please refer to Journal-level guidance for any specific requirements.

\section*{Declarations}

Some journals require declarations to be submitted in a standardised format. Please check the Instructions for Authors of the journal to which you are submitting to see if you need to complete this section. If yes, your manuscript must contain the following sections under the heading `Declarations':

\begin{itemize}
\item Funding
\item Conflict of interest/Competing interests (check journal-specific guidelines for which heading to use)
\item Ethics approval 
\item Consent to participate
\item Consent for publication
\item Availability of data and materials
\item Code availability 
\item Authors' contributions
\end{itemize}

\noindent
If any of the sections are not relevant to your manuscript, please include the heading and write `Not applicable' for that section. 

\bigskip
\begin{flushleft}%
Editorial Policies for:

\bigskip\noindent
Springer journals and proceedings: \url{https://www.springer.com/gp/editorial-policies}

\bigskip\noindent
Nature Portfolio journals: \url{https://www.nature.com/nature-research/editorial-policies}

\bigskip\noindent
\textit{Scientific Reports}: \url{https://www.nature.com/srep/journal-policies/editorial-policies}

\bigskip\noindent
BMC journals: \url{https://www.biomedcentral.com/getpublished/editorial-policies}
\end{flushleft}

\begin{appendices}

\section{Section title of first appendix}\label{secA1}

An appendix contains supplementary information that is not an essential part of the text itself but which may be helpful in providing a more comprehensive understanding of the research problem or it is information that is too cumbersome to be included in the body of the paper.




\end{appendices}


\bibliography{sn-bibliography} % common bib file

\end{document}